\documentstyle[12pt]{article}

\def\be{\begin{equation}}
\def\te{\end{equation}}
\def\bea{\begin{eqnarray}}

\def\tea{\end{eqnarray}}

\makeatletter 
\@addtoreset{equation}{section}
\makeatother  

\begin{document}

\title{Stochastic Behavior of Effective Field Theories Across Threshold}
\author{E. Calzetta \\
{\small IAFE and FCEN, University of Buenos Aires, Argentina}\\
B. L. Hu \\
{\small Department of Physics, University of Maryland, College Park, MD
20742, USA}}
\date{{\small {\it (hep-th/9603164, UMDPP 96-90, March 20, 1996)}}}
\maketitle

\begin{abstract}
We explore how the existence of a field with a heavy mass influences the low
energy dynamics of a quantum field with a light mass by expounding the
stochastic characters of their interactions which take on the form of 
fluctuations in the number of (heavy field) particles created at the
threshold, and  dissipation in the dynamics of the light fields,
arising from the backreaction of produced heavy particles. We claim that the
stochastic nature of effective field theories is intrinsic, in that
dissipation and fluctuations are present both above and below the threshold.
Stochasticity builds up exponentially quickly as the heavy threshold is
approached from below, becoming dominant once the threshold is crossed. But
it also exists below the threshold and is in principle detectable, albeit
strongly suppressed at low energies. The results derived here can be used to
give a quantitative definition of the `effectiveness' of a theory in terms
of the relative weight of the deterministic versus the stochastic behavior
at different energy scales.
\end{abstract}


\newpage

\section{Introduction}

The goal of this paper is to explore how the existence of a field with a
heavy mass (a heavy sector, or heavy field, in short) influences the low
energy dynamics of a quantum field with light mass (light sector, or light
field). It is a well known result from effective field theory \cite
{eft,weinberg,eft2} that at low energies the heavy fields effectively
decouple. This means that it is possible to describe the low energy
(infrared) physics through an effective action of the light fields, whereby
no explicit reference to heavy fields is made. \footnote{%
It should be clear from the discussion below that we regard the effective
theory as a description of the actual physics accessible to an observer at a
given energy scale including the effect of the higher mass sector, rather
than as a formal construct obtained from the full theory by application of
some approximation scheme such as the Schwinger-DeWitt proper-time
quasilocal \cite{SchDeW} or the Heisenberg-Euler \cite{HeiEul}
inverse-mass expansion of propagators.
The nonperturbative effect we discuss here cannot be obtained by these
approximations.} This description breaks down as the energy gets close to
the heavy particle mass threshold, where particle creation of the heavy
field begins to get significant. Generally, the breakdown of the effective
light theory is described in terms of the loss of predictive power of the
theory, resulting from a proliferation of increasingly nonlocal
interactions. We shall propose a new way to look at the threshold behavior
of the light theory, based on the manifestation of stochastic features which
we believe are intrinsic to effective theories \cite{HuPhysica,Banff}.

This loss of predictability follows from the fact that the light field
interacts in a complex way with quantum fluctuations of the heavy field. The
stochastic characters of these interactions take on the form of {\it %
fluctuations} in the number of (heavy field) particles created at the
threshold, and {\it dissipation} in the dynamics of the light fields,
arising from the backreaction of produced heavy particles. The stochastic
properties of effective field theory arise from particle creation, and the
two processes it engenders, i.e., dissipation and noise, are related by the
fluctuation-dissipation relation \cite{fdt}.

The appearance of stochastic behavior like dissipation and noise is
predicated upon the actual observational context which defines the system,
and its interaction with the unobserved or unobservable variables which make
up the environment. In fact, for observers in the limited range of validity
of the system (say, at low energy), the existence of the environment (say,
the heavy sector) can sometimes only be indirectly deduced by the modified
behavior of the system, rendered by such restrictions.

Dissipation becomes obvious above the heavy mass threshold, where the light
field self energy becomes imaginary (in agreement with the optical theorem).
As we shall show below, a proper analysis of the effective light theory
shows that a dissipative theory must also be stochastic at some level. Our
claim, based on the results of this paper, is that the stochastic nature of
effective field theories is intrinsic, in that dissipation and fluctuations
are present both above and below threshold. Stochasticity builds up
exponentially quickly as the heavy threshold is approached, becoming
dominant once the threshold is crossed. But it also exists below threshold,
albeit strongly suppressed at low energy. This is in contradistinction to
the conventional belief that such behavior changes discontinuously on
threshold crossing. Using the expressions we derive here, one can
quantitatively define the `effectiveness' of a theory in terms of the
relative weight of the deterministic versus the stochastic behavior at
different energy scales.

The presence of stochastic effects shows that the physics of the light
fields is different in an effective theory, in small but important ways,
from what would follow if the light action were fundamental, even at scales
below the heavy threshold. The difference lies in the phenomena of
dissipation and fluctuation generation, which are present in effective
theories but absent in fundamental ones. On a more speculative level, we may
argue that there is no real `fundamental' theory in nature \cite{HuSpain}. A
theory only appears to be fundamental at low energy in ignorance of the
presence of other heavy constituents in nature because the stochastic
components generated from such interactions are suppressed. At higher
energies such features become more important and the presence of the heavy
sector becomes more apparent. Thus, the magnitude of noise and dissipation
can serve as a measure of the degree of resolution of the means of
observation compared to the intrinsic mass or energy scales of the more
complete theory. This viewpoint is a natural consequence from regarding an
effective theory as an open system \cite{HuPhysica,Banff}, which is what we
used earlier in the analysis of the statistical mechanical properties of
particles and quantum fields \cite{HPZ,HM2,dch,cddn} and semiclassical
gravity \cite{nfsg,HM3,fdrsc,qfsf}.

Recent investigation of the statistical mechanical aspect of gravitational
systems and quantum fields began with the work of Bekenstein \cite{Bek73}
and Hawking \cite{Haw75} on black hole entropy. Penrose's proposal of
gravitational entropy and the Weyl curvature conjecture \cite{Penrose} was
analyzed in the context of backreaction of particle creation by one of us. 
\cite{hu82}. Entropy of quantum fields associated with particle creation was
discussed in \cite{hupav} and \cite{hukan} 
(see also \cite{ccs}). The concept of field entropy was further explored in 
\cite{bmp,GasGio} and more recent works. Entropy of interacting fields
defined by the truncation of a BBGKY hierarchy and the factorization of
higher order correlation functions was proposed in \cite{CH88}. Noise,
decoherence, fluctuations and dissipation in this scheme were discussed in 
\cite{dch,cddn}. A common assumption in quantum theories of structure
formation, i.e., quantum correlation functions directly go over to their
classical counterparts, was shown to be incorrect \cite{qfsf} when 
the stochastic properties of quantum fields are carefully considered. As our
present analysis further demonstrates, only those modes of the light field
which are dynamically entangled with the quantum heavy fields can partake of
the process of decoherence and quantum to classical transition to acquire a
stochastic character. The relationship between dissipation and stochasticity
has been further discussed in the context of decoherence and quantum to
classical transition. Calzetta and Mazzitelli pointed out the connection
between particle creation and decoherence \cite{calmaz}. Paz and Sinha
showed that a decohered field must of necessity possess traits of randomness 
\cite{pazsin}. The stochastic aspects of classical theories emerging from
quantum mechanics are discussed at length in an important paper by Hartle
and Gell-Mann \cite{hargel}.(See also \cite{Tsukuba}.)

Our inquiry into the stochastic nature of effective field theory thus
compels us to adopt a new and more complex viewpoint of quantum field
theory, incorporating the statistical mechanical properties of quantum
fields. This means that we are more interested in the {\it causal development%
} of quantum fields than the traditional scattering or transition amplitude
aspects. For this we need the in-in (or closed time path CTP, or
Schwinger-Keldysh) \cite{ctp} rather than the in-out (or Schwinger-DeWitt) 
\cite{SchDeW} formulation. We will use the related influence functional
(Feynman-Vernon) \cite{if} formalism to extract the stochastic features of
effective field theory. In addition, we will need to probe into the {\it %
nonperturbative effects}. By perturbative, we refer here specifically to
expansions in the coupling constants of fields, rather than loop expansion,
or adiabatic approximation. Non perturbativew calculations are exemplified
by Schwinger's original derivation of particle creation in a constant
electric field \cite{Schwinger51}. For noninteracting fields in curved
spacetimes with regions where a vacuum for a field theory can be defined
(asymptotically flat, or statically- bounded evolution), Parker \cite{Par69}
and Zel'dovich's \cite{Zel70} treatment of cosmological particle creation
and Hawking's \cite{Haw75} derivation of black hole radiance by means of
Bogolubov transformations are nonperturbative, even though for more general
situations where a well-defined global vacuum is lacking \cite{Ful73}, one
may need to appeal to approximate or perturbative concepts such as adiabatic
vacuum \cite{ParFul,ZelSta}. For fields propagating in nontrivial spacetimes
(such as the nonconformally-flat spacetimes of the Bianchi Universes \cite
{harhu,CH87}), or for interacting fields (e.g. \cite{HZ88,CH89}) one has to
appeal to perturbative expansion of the interaction or coupling parameter
(such as the $\lambda $ in a $\phi ^4$ theory or the anisotropy). These
nonperturbative effects may be quantitatively significant in the proper
environment, such as during the reheating era in inflationary cosmology \cite
{StyPhD,Paz90,RHS,stabra,Boyan}.

The main result of this paper is that a light field plane wave is always
followed by a stochastic, slowly varying light `echo'. This `echo' is
produced by the back reaction of heavy particles created from the seed light
wave, and it is a nonperturbative effect. The amplitude and growth rate of
the echo increases exponentially as the frequency of the seed wave
approaches the heavy scales. We may understand this as a `diamagnetic'
effect (in contrast to a paramagnetic effect), since it involves two steps:
First the polarization of the vacuum by the light seed wave, and then the
coupling of the appropriate light modes to the polarized vacuum.

If the light field self-interacts, this effect shall be masked by the
corresponding one originating from quantum fluctuations of the light field.
In principle, these two effects could be disentangled by recourse to their
different scale dependence. In any case, the presence of both effects
underlies the fact that any field theory used in practice for the
description of real physical systems is an effective theory. This heuristic
observation may be put on a rigorous footing by casting a field theory as a
theory of a background field interacting with a hierarchy of Wightman
functions; the theory becomes effective when this hierarchy is truncated,
either explicitly through some approximation scheme, or implicitly by the
limited accuracy of specified observations \cite{dch,cddn}.

To connect with earlier theoretical work (developed mainly in the 70's and
80's) we shall begin in Sec. 2 with a discussion of conventional effective
field theories \cite{eft,eft2}. We shall show how the features of light
physics we want to develop appear already in the conventional (in-out)
formulation, albeit in a somewhat obscure form. We shall then in Sec. 2.2 go
over to the more suitable causal (in-in) formulation of quantum field
theory. In Sec. 2.3 we show how the stochastic characters of an effective
field theory can be identified from the in-in effective action with the aid
of the Feynman- Vernon formalism, and how both dissipation and fluctuations
can be related to particle creation above threshold. In Sec. 3 we show how
such features remain in the below threshold regime and derive the new
effects which can in principle be used to discern an effective theory from a
fundamental theory. The Appendices contain the details of the derivation of
the heavy field quantization in the background of a light plane wave.

\section{Perturbative effective field theory}

As could be expected, the problems with conventional light effective
theories are most acute if we attempt to implement them in the above
threshold regime; in this range, dissipation and noise appear even at the
perturbative level. We shall take advantage of this fact to introduce the
main concepts of dressed field, dissipation and noise kernels, etc, in the
familiar setting of one-loop Feynman graphs. We shall then proceed in the
next section to discuss these effects in the physically more relevant,
below-threshold regime.

\subsection{Effective Theory of the Light Particles}

To simplify the technical burden, we shall work on a toy model of quantum
field theory consisting of two real scalar fields $\phi $ and $\Phi $. We
have used this model to treat the dissipation of quantum fields via particle
creation by the CTP method before \cite{CH87,CH89}. For more general models,
we refer the reader to \cite{HPZ}. Similar consideration of the dissipative
and noise properties of two field interactions can be found in \cite
{StyPhD,Paz90,RHS,Boyan,cgea,SinHu,HuBelgium}.

The classical action is given by

\begin{equation}
S=S_l+S_H+S_{lH}
\end{equation}

\begin{equation}
S_l=\int d^4x\;\left( -\frac 12\right) \left[ \partial \phi \partial \phi
+m^2\phi ^2-g\left\langle \Phi ^2\right\rangle _0\phi \right]
\end{equation}

\begin{equation}
S_H=\int d^4x\;\left(- \frac{1}2\right) \left[ \partial \Phi \partial \Phi
+M^2\Phi ^2\right]
\end{equation}

\begin{equation}
S_{lH}=\int d^4x\;\left( -\frac 12\right) g\phi \Phi ^2
\end{equation}
where the subscripts $l,H$ are used to denote the light and heavy fields. We
use signature -+++, and ignore terms necessary for renormalization purposes
(other than the term linear in the $\phi $ field) Here, $\left\langle \Phi
^2\right\rangle _0$ stands for the vacuum expectation value in the absence
of background field, i.e., $\left\langle \Phi^2\right\rangle _{\phi=0}$. We
also assume $m\ll M$.

The effective theory of light particles (we may henceforth call it light
effective theory) is defined by the action functional

\begin{equation}
S_{eff}=S_l+\delta S
\end{equation}
where

\begin{equation}
\delta S\left[ \phi \right] =-i\ln \int D\Phi \;e^{i\left\{ S_H\left( \Phi
\right) +S_{lH}\left( \phi ,\Phi \right) \right\} }  \label{deltas}
\end{equation}
Formally, $\delta S$ is the effective action for the heavy fields
propagating on the light background field (considered as an external field),
evaluated at the vacuum expectation value \cite{jim}. In our case, this VEV
vanishes by symmetry, so we shall not mention it explicitly. Perturbatively, 
$\delta S$ is the sum of all 1PI vacuum bubbles, with light field insertions
but only heavy internal lines.

To second order in the coupling constant $g$, we find

\begin{equation}
\delta S\left[ \phi \right] =\left( \frac{-g}2\right) \int d^4x\;\phi
(x)\Delta _F\left( x,x\right) +\left( \frac{ig^2}4\right) \int
d^4xd^4x^{\prime }\phi (x)\phi (x^{\prime })\Delta _F^2\left( x,x^{\prime
}\right)
\end{equation}
where

\begin{equation}
\Delta _F\left( x,x^{\prime }\right) =\frac{<OUT\left| T(\Phi (x)\Phi
(x^{\prime }))\right| IN>}{<OUT\mid IN>}=(-i)\int \frac{d^4k}{\left( 2\pi
\right) ^4}e^{ik(x-x^{\prime })}\frac 1{k^2+M^2-i\varepsilon }
\end{equation}
is the Feynman propagator for the heavy particles. The linear term is
cancelled by an appropriate counterterm. For the quadratic term, we find

\begin{equation}
\Delta _F^2\left( x,x^{\prime }\right) =\int \frac{d^4k}{\left( 2\pi \right)
^4}e^{ik(x-x^{\prime })}\left\{ (i)(k^2+m^2)\int_{4M^2}^\infty \frac{ds}{%
(s-m^2)}\frac{h(s)}{(k^2+s-i\varepsilon )}\right\}  \label{quad}
\end{equation}
where we have performed the necessary subtractions to insure that $m^2$
remains the physical mass of the light field, and

\begin{equation}
h(s)=\frac 1{(4\pi )^2}\sqrt{1-\frac{4M^2}s}  \label{hs}
\end{equation}

An effective light theory deals with the $k^2\rightarrow 0$ limit. We can
obtain a formal expression for the effective action by expanding Eq. (\ref
{quad}) in inverse powers of the heavy mass; this expansion is analogous to
the Heisenberg-Euler lagrangian for the electromagnetic field \cite{HeiEul}. 
At any
finite order, we obtain a higher derivative theory \cite{lucho}.
Such approximation will not show dissipation nor fluctuations in the
light field.

In this limit, $\delta S$ is analytic in $k$ and real. However,
this ceases to be the case as soon as we cross the heavy
particle threshold $4M^2$. Above the threshold, $\delta S$ is neither
analytical nor real. In this regime a light effective theory is not only
cumbersome because of the proliferation of nonlocal terms, rather, the whole
concept of an effective action breaks down.

A striking feature of the light action, if we insist on taking it seriously
above threshold, is that it leads to complex and noncausal equations of
motion. This follows from the IN OUT boundary conditions built in the path
integral Eq. (\ref{deltas}). The imaginary part of the effective action is
related to the imaginary part of the Feynman graph. Because of the optical
theorem, we know this in turn is related to pair creation from the light
particles. Therefore, a complex action would give rise to dissipative terms
in the equation of motion of the light field, and the fluctuations in the
particle creation would measure the breakdown of the (low energy) effective
theory. The unitarity of the full quantum theory is broken in the effective
theory. However, the chosen IN-OUT boundary conditions obscure this fact,
since they make it hard to discern the arrow of time arising from
dissipation \cite{harhu}. For this and other reasons, one should use
the causal, IN-IN, boundary conditions, as introduced by Schwinger
\cite{ctp,CH87,CH89}.

\subsection{Causal effective field theory}

Let us derive the causal and real equations of motion which describe the
evolution of physical perturbations of the light field. This is achieved by
doubling the degrees of freedom to two fields $\phi ^{+,-}$, or rather, by
assuming the field is actually defined on a closed time path. The equations
of motion are found by taking the variation with respect to $\phi ^+$ of the
CTP action functional

\begin{equation}
S_{eff}^{CTP}=S_l\left[ \phi ^{+}\right] -S_l\left[ \phi ^{-}\right] +\delta
S^{CTP}\left[ \phi ^{+},\phi ^{-}\right]
\end{equation}
where \cite{ctp,CH87}

\begin{equation}
\delta S^{CTP}\left[ \phi ^+,\phi ^-\right] =-i\ln \int D\Phi ^+D\Phi ^-\;
e^{i\left\{ S_H\left( \Phi ^+\right) -S_H\left( \Phi ^+\right) +S_{lH}\left(
\phi ^+,\Phi ^+\right) -S_{lH}\left( \phi ^-,\Phi ^-\right) \right\} }
\label{dsctp}
\end{equation}

The quadratic terms in the effective action, to second order in the coupling
constant, are

\begin{equation}
\frac{ig^2}4\int d^4xd^4x^{\prime }\left\{ \phi ^{+}(x)\phi ^{+}(x^{\prime
})\Delta _F^2\left( x,x^{\prime }\right) -2\phi ^{+}(x)\phi ^{-}(x^{\prime
})\Delta _{-}^2\left( x,x^{\prime }\right) +\phi ^{-}(x)\phi ^{-}(x^{\prime
})\Delta _D^2\left( x,x^{\prime }\right) \right\}  \label{quadterms}
\end{equation}
where the propagators are the expectation values taken with respect to the
IN vacuum defined by

\begin{equation}
\Delta _F\left( x,x^{\prime }\right) =<IN\left| T(\Phi (x)\Phi (x^{\prime
}))\right| IN>=(-i)\int \frac{d^4k}{\left( 2\pi \right) ^4}e^{ik(x-x^{\prime
})}\frac 1{k^2+M^2-i\varepsilon }
\end{equation}

\begin{equation}
\Delta _D\left( x,x^{\prime }\right) =<IN\left| \widetilde{T}(\Phi (x)\Phi
(x^{\prime }))\right| IN>=(i)\int \frac{d^4k}{\left( 2\pi \right) ^4}%
e^{ik(x-x^{\prime })}\frac 1{k^2+M^2+i\varepsilon }
\end{equation}

\begin{equation}
\Delta _{-}\left( x,x^{\prime }\right) =<IN\left| \Phi (x^{\prime })\Phi
(x)\right| IN>=(2\pi )\int \frac{d^4k}{\left( 2\pi \right) ^4}%
e^{ik(x-x^{\prime })}\delta \left( k^2+M^2\right) \theta \left( -k^0\right)
\end{equation}

The equations of motion becomes (but see below!)

\begin{equation}
\left( -\Box +m^2\right) \phi (x)+g^2\int d^4x^{\prime }\;D\left(
x,x^{\prime }\right) \phi (x^{\prime })=0  \label{leq}
\end{equation}
where

\begin{equation}
\;D\left( x,x^{\prime }\right) =\frac i2\left[ \Delta _F^2\left( x,x^{\prime
}\right) -\Delta _{-}^2\left( x,x^{\prime }\right) \right]
\end{equation}

Since from the definitions

\begin{equation}
\Delta _{-}\left( x,x^{\prime }\right) =\Delta _F\left( x,x^{\prime }\right)
~~ if ~~ t^{\prime }>t,
\end{equation}
while

\begin{equation}
\Delta _{-}\left( x,x^{\prime }\right) =\Delta _F^{*}\left( x,x^{\prime
}\right) ~~if~~t>t^{\prime },
\end{equation}
it is obvious that Eq. (\ref{leq}) is real and causal. Explicitly

\begin{equation}
\;D\left( x,x^{\prime }\right) =\int \frac{d^4k}{\left( 2\pi \right) ^4}%
e^{ik(x-x^{\prime })}\left\{ (\frac{-1}2)(k^2+m^2)\int_{4M^2}^\infty \frac{ds%
}{(s-m^2)}\frac{h(s)}{[(k+i\varepsilon )^2+s]}\right\}
\end{equation}
with the same $h$ as in Eq. (\ref{hs}), and $(k+i\varepsilon
)^2=-(k^0+i\varepsilon )^2+\vec k^2$, carrying the causal boundary
conditions.

The light field $\phi $ described by the wave equation Eq. (\ref{leq}) is
clearly no longer the classical light field, but is now dressed through the
interaction with the quantum fluctuations of the heavy field. A different
approach to the dynamics clarifies this point. The Heisenberg equations of
motion for the light field are

\begin{equation}
\left( -\Box +m^2\right) \phi (x)+(\frac g2)\left[ \Phi ^2-<\Phi
^2>_0\right] =0  \label{hleq}
\end{equation}
where we have substracted the expectation value of $\Phi ^2$, computed at
vanishing light fields, to make $\phi =0$ the true light vacuum. Comparing (%
\ref{leq}) and (\ref{hleq}), we see that the former amounts to the
approximation

\begin{equation}
\left[ \Phi ^2-<\Phi ^2>_0\right] \approx 2g\int d^4x^{\prime }\;D\left(
x,x^{\prime }\right) \phi (x^{\prime })
\end{equation}
On the other hand, a direct calculation shows that

\begin{equation}
2g\;D\left( x,x^{\prime }\right) \equiv \frac{\delta <\Phi ^2>(x)}{\delta
\phi (x^{\prime })}\mid _{\phi =0}
\end{equation}
so that, within the present accuracy,

\begin{equation}
2g\int d^4x^{\prime }\;D\left( x,x^{\prime }\right) \phi (x^{\prime })\simeq
<\Phi ^2>_\phi -<\Phi ^2>_0 
\end{equation}
Here, $<\Phi ^2>_\phi $ stands for the vacuum expectation value evaluated
with respect to the background of a non-zero light field $\phi $. So in this
approximation, the q-number $\Phi ^2$ in the Heisenberg equation of motion
is substituted by its expectation value, computed as a causal functional of
the light background. (In quantum open systems language, the heavy field is
said to be `slaved' to the light one \cite{cddn}).

Now the expectation value does not give a faithful description of the action
of the heavy fields on the light ones. Indeed, this approximation will
become inaccurate as heavy pair creation accumulates and the actual state of
the heavy field deviates, in unpredictable ways, from the vacuum. It is in
this way that an arrow of time appears in the theory. Pair creation brings
forth dissipation, and the fluctuations in back reaction associated to
fluctuations in particle number manifest as noise, as we shall presently
show.

\subsection{Above Threshold: Fluctuations and Dissipation}

The expectation value alone does not capture the full effect of the heavy
fields on the light ones. Investigations in recent years show that
theoretical consistency demands that an extra stochastic source term $\xi
(x) $ should be present in (\ref{leq}), as 
\begin{equation}
\left( -\Box +m^2\right) \phi (x)+g^2\int d^4x^{\prime }\;D\left(
x,x^{\prime }\right) \phi (x^{\prime })=g\xi (x),  \label{lanleq}
\end{equation}
which takes into account the fluctuations of the heavy fields, namely,

\begin{equation}
\xi(x)\sim \left( \frac 12\right) \left[ \Phi ^2-<\Phi ^2>_\phi \right] (x)
\end{equation}
The external source vanishes on average, but has a $rms$ value

\begin{equation}
\left\langle \xi(x)\xi(x^{\prime })\right\rangle \equiv N(x,x^{\prime
})=\left( \frac 18\right) \left[ \left\langle \left\{ \Phi ^2(x),\Phi
^2(x^{\prime })\right\} \right\rangle _0-2<\Phi ^2>_0^2\right]
\end{equation}
Or, explicitly,

\begin{equation}
N(x,x^{\prime })=\int \frac{d^4k}{\left( 2\pi \right) ^4}e^{ik(x-x^{\prime
})}\left( \frac \pi 2\right) h(-k^2)  \label{noiseker}
\end{equation}

Following Feynman and Vernon \cite{if}, as we have done in related problems 
\cite{qfsf}, we can show that a Langevin type equation Eq. (\ref{lanleq}) is
properly derived from the CTP effective action, rather than the more
familiar deterministic equation Eq. (\ref{leq}).

To this end, let us first replace the field variables $\phi ^{+, -}$ by the
average and difference variables

\begin{equation}
\left[ \phi \right] =\phi ^+ - \phi ^-, ~~ \left\{ \phi \right\} =\phi ^+ +
\phi ^-
\end{equation}
With the identity

\begin{equation}
S^{CTP}_{eff}\left[ \left\{ \phi \right\} ,\left[ \phi \right] =0\right]
\equiv 0
\end{equation}
it follows that the equation of motion is

\begin{equation}
\frac{\delta S^{CTP}_{eff}}{\delta \left[ \phi \right] }\left[ \left\{ \phi
\right\} =\phi ,\left[ \phi \right] =0\right] =0
\end{equation}

The quadratic terms in the effective action, Eq. (\ref{quadterms}), may be
written as

\begin{equation}
\frac{g^2}2\int d^4xd^4x^{\prime }\left\{ -\left[ \phi (x)\right] D\left(
x,x^{\prime }\right) \left\{ \phi (x^{\prime })\right\} +i\left[ \phi
(x)\right] N\left( x,x^{\prime }\right) \left[ \phi (x^{\prime })\right]
\right\}
\end{equation}
It may seem that the noise kernel $N$ does not contribute to the equations
of motion. However, by virtue of the identity

\begin{equation}
\exp \left\{ -\frac{g^2}2\int d^4xd^4x^{\prime }\left[ \phi (x)\right]
N\left( x,x^{\prime }\right) \left[ \phi (x^{\prime })\right] \right\}
\equiv \int D \xi \;P \left[ \xi \right] \exp \left\{ -ig\int d^4x\;\xi(x)
\left[\phi (x)\right] \right\}
\end{equation}
for some probability density $P$, with

\begin{equation}
\left\langle \xi (x)\xi (x^{\prime })\right\rangle \equiv N(x,x^{\prime })
\end{equation}
we may substitute the quadratic term in the light effective action by
coupling the field to a stochastic source whose auto-correlation is given by
the noise kernel $N$. The fact that both dissipation and noise kernels can
be expressed in terms of the same function $h$ in this example is the origin
of the fluctuation-dissipation theorem.

It should be realized that the inclusion of a stochastic source in Eq. (\ref
{lanleq}) is not merely an improvement over the earlier Eq. (\ref{leq}), but
it is required for consistency of the light effective theory. \footnote{%
The dressed light field is {\it prima facie} a quantum field, and the
stochastic driving force from the environment is also quantum in nature.
Since light and heavy modes are dynamically entangled, interference
effects are abound, but are extremely hard to display for observations carried
out at low energy. A heavy sector serving as an environment to the
light sector can  decohere it, and induce a quantum to classical
transition. After decoherence the open system variables obey an effectively
classical equation of motion, but driven by stochastic source terms, such as
in a Langevin equation. Indeed, the amount of noise in this open system is a
direct measure of the degree of entanglement with the unobserved sector.
(This delicate borderline between classical and quantum physics is a general
feature of quantum noisy systems \cite{gardiner}.) The issue of decoherence
is an important one lying at the foundation of quantum mechanics and has
been studied by many people in recent years \cite{envdec,conhis}. We have also
discussed this issue for model field theories \cite{nfsg,qfsf}.
By the  same reasoning, we can assume
safely for our considerations here that the light field has been decohered
and behaves like a classical stochastic field.}

We can see this, for example, by consideration of the light field
fluctuations. These are described by the Hadamard kernel

\begin{equation}
G_1(x,x^{\prime })=\left\langle IN\left| \left\{ \phi (x),\phi (x^{\prime
})\right\} \right| IN\right\rangle 
\end{equation}
The Fourier transform $G_1(k)$ is related to that of 
the expectation value of the commutator of two fields $G$ by the zero
temperature KMS formula \cite{kms}

\begin{equation}
G_1(k)={\rm sign}(k^0)G(k) 
\end{equation}

If we assume canonical equal-time commutation relations, there is a simple
relationship between $G$ and $G_{ret}$ 

\begin{equation}
G_{ret}(x,x^{\prime })=(-i)G(x,x^{\prime })\theta (x^0-x^{0^{\prime }}) 
\end{equation}
which translates into

\begin{equation}
G(k)=2\;{\rm Im\;}G_{ret}(k) 
\end{equation}

The retarded propagator is simply the inverse of Eq. (\ref{leq}) with causal
boundary conditions. It has a pole at $-k^2=m^2$ and a branch cut from $%
-k^2=4M^2$ on. Therefore

\begin{equation}
G_{ret}(k)=\frac B{[(k+i\varepsilon )^2+m^2]}+\left( \frac{g^2}2\right)
\int_{4M^2}^\infty ds\;\frac{h(s)\left| G_{ret}(s)\right| ^2}{%
[(k+i\varepsilon )^2+s]} 
\end{equation}
where $B$ is the residue at the pole, and $G_{ret}(s)$ stands for the
propagator evaluated on the $-k^2=s$ shell. We conclude

\begin{equation}
G(k)=2\pi \left\{ B\delta \left( k^2+m^2\right) +\left( \frac{g^2}2\right)
h(-k^2)\left| G_{ret}(k)\right| ^2\theta \left( -k^2-4M^2\right) \right\} 
{\rm sign}(k^0) 
\end{equation}

\begin{equation}
G_1(k)=2\pi \left\{ B\delta \left( k^2+m^2\right) +\left( \frac{g^2}2\right)
h(-k^2)\left| G_{ret}(k)\right| ^2\theta \left( -k^2-4M^2\right) \right\}
\label{lff}
\end{equation}

But this result is inconsistent with Eq. (\ref{leq}): since modes above
threshold are damped, they couldn't possibly sustain a time translation
invariant auto-correlation such as Eq. (\ref{lff}). However, addition of the
stochastic source $g\xi $ removes the contradiction: we can identify the
first term in Eq. (\ref{lff}) as the `natural' light quantum fluctuations,
and the second as the fluctuations induced by the action of the external
source, in agreement with Eq. (\ref{noiseker}). The stochastic source feeds
onto the light field precisely the amount of fluctuation necessary to keep
the noise level, as required by the fluctuation-dissipation relation.

Are these quantum or classical fluctuations? The calculation above relied on
ordinary quantum mechanical rules, such as the KMS theorem, so one would
answer they are quantum fluctuations. But this is the point of view of an
observer who is aware of the existence of the heavy field. Since the high
momentum modes are entangled with the heavy field and become correlated
through particle creation, only such an observer could effect interference
between modes above the threshold. For observers who cannot operate on the
heavy field, the interference of these modes are not observable, and his
answer could be that these are classical fluctuations. As in many other
situations in quantum physics, questions like this can have different
answers depending on the specific observational context.

Our perturbative treatment so far suggests that noise and dissipation only
turn on above threshold. We now wish to show that they are indeed present
below the threshold, albeit exponentially suppressed.

\section{Nonperturbative effective theory}

The analysis of the previous Section highlighted the main elements of the
light field theory, namely, the dressing of the light field by the heavy
quantum fluctuations, the onset of dissipative processes, and the
decoherence and noise generation thereof. However, this analysis is based on
a hypothetical light field with momenta above the heavy threshold, a regime
where light effective theory would be of theoretical rather than practical
(observational) interest. What makes the ongoing discussion relevant is that
the same phenomena are actually occurring at the light scales, albeit
strongly suppressed. They are not revealed in a perturbative theory. In this
Section, we shall describe some of the most conspicuous manifestations of
noise and dissipation in the infrared regime.

To simplify the calculation, we shall assume the existence of a seed
classical light background field, in the form of a monochromatic plane wave.
It may arise through the action of some external agent, or as an outcome of
the previous history of the system. We also assume that the interaction
between the light background and the heavy quantum field is adiabatically
switched off in the past, so that there is a well defined IN vacuum for the
heavy fields, and that no substantial particle creation occurs prior to a
given time (conventionally taken as $t=0$), so that at this time the heavy
field is still in the IN vacuum state. Our aim is to compute the
amplification of the heavy quantum fluctuations due to parametric resonance,
and the light fluctuations arising from the back-reaction thereof. 

This situation actually arises in many cases of interest, like the
background gravitational field in the early Universe interacting with
quantum matter fields. In this case, detailed studies show that indeed the
gravitational field is prone to decohere earlier than the matter fields, so
the classical - quantum distinction is unambiguous \cite{pazsin}. In the
case of multiparticle production in heavy ion collisions, for example,
particle currents are applied externally, while the gauge fields take the
role of the `irrelevant' heavy fields \cite{hic} which are coarse-grained.
If we study the generation of a cosmic background magnetic field, on the
other hand, a seed magnetic field comes from the past (for example, through
amplification of vacuum fluctuations during inflation) and is further
amplified through interaction with charged particles in the radiation era 
\cite{tw}. To give yet another example, we could model a laser as a light
field (the electromagnetic field in a resonant cavity) interacting with
heavy fields (the creation operators for the gas in the cavity, in different
possible internal states). Then the seed is the externally provided pumping 
\cite{korenman}.

While the situation we shall discuss is at best a toy model for these
relevant systems, it will allow us to show in detail how the back reaction
of the heavy field on the light one leads to the onset of a distinct,
inhomogenous, stochastic structure, whose amplitude, growth, and coarsening
rates depend exponentially on the ratio of the light to the heavy scales.
Thus the light theory will have a stochastic character, even for observers
confined to infrared phenomenology.

\subsection{Nonperturbative equations of motion}


Let us return to the fundamental definitions

\begin{equation}
S_{eff}^{CTP}=S_l\left[ \phi ^{+}\right] -S_l\left[ \phi ^{-}\right] +\delta
S^{CTP}\left[ \phi ^{+},\phi ^{-}\right]
\end{equation}

\begin{equation}
\delta S^{CTP}\left[ \phi ^{+},\phi ^{-}\right] =-i\ln \int D\Phi ^{+}D\Phi
^{-}\;e^{i\left\{ S_H\left( \Phi ^{+}\right) -S_H\left( \Phi ^{+}\right)
+S_{lH}\left( \phi ^{+},\Phi ^{+}\right) -S_{lH}\left( \phi ^{-},\Phi
^{-}\right) \right\} }
\end{equation}

We shall now attemp a nonperturbative evaluation of this path integral.
Using the sum and difference field variables

\begin{equation}
\left[ \phi \right] =\phi ^{+}-\phi ^{-},~~\left\{ \phi \right\} =\phi
^{+}+\phi ^{-}
\end{equation}
we can extract the `deterministic' part as

\begin{equation}
\delta S^{CTP}\left[ \phi ^{+},\phi ^{-}\right] =\left( \frac{-g}2\right)
\int d^4x<\Phi ^2>_{\left\{ \phi \right\} }(x)\left[ \phi \right] +\Delta
S\left[ \left\{ \phi \right\} ,\left[ \phi \right] \right] \qquad \; 
\end{equation}
where as defined in the previous section, the subscript $\{\phi\}$ denotes
averaging with respect to the $\{\phi\}$ field. We perform a functional
Fourier transform

\begin{equation}
\exp \left\{ i\Delta S\left[ \left\{ \phi \right\} ,\left[ \phi \right]
\right] \right\} =\int D\xi \;e^{ig\int \xi \left[ \phi \right] }P\left[ \xi
,\left\{ \phi \right\} \right] 
\end{equation}

Observe that

\begin{equation}
\left\langle \xi (x)\right\rangle =0 
\end{equation}

\begin{equation}
\left\langle \xi (x)\xi (x^{\prime })\right\rangle \equiv N(x,x^{\prime
})=\left( \frac 18\right) \left[ \left\langle \left\{ \Phi ^2(x),\Phi
^2(x^{\prime })\right\} \right\rangle _{\left\{ \phi \right\} }-2<\Phi
^2>_{\left\{ \phi \right\} }(x)<\Phi ^2>_{\left\{ \phi \right\} }(x^{\prime
})\right]
\end{equation}
where

\begin{equation}
\left\langle f\right\rangle \equiv \int D\xi \;f\;P\left[ \xi ,\left\{ \phi
\right\} \right] 
\end{equation}
This is to be contrasted with the result in the perturbative treatment
Eq. (\ref{noiseker}).

The functional $P\left[ \xi ,\left\{ \phi \right\} \right] $ must be real
(as follows from $\Delta S\left[ \left\{ \phi \right\} ,-\left[ \phi \right]
\right] =-\Delta S\left[ \left\{ \phi \right\} ,\left[ \phi \right] \right]
^{*}$) and it is nonnegative to 1 loop approximation. We may think of it as
a functional Wigner transform of the effective action \cite{salman}, and
thereby as a probability density `for all practical purposes'. Observe that $%
P$ will not be Gaussian in general. In our concrete application,
nevertheless, the effective action is 1-loop exact, so the identification of 
$P$ as a Gaussian probability density poses no difficulty.

We conclude that the correct, nonperturbative effective equation of motion
for the light fields reads

\begin{equation}
\left( -\Box +m^2\right) \phi (x)+(\frac g2)\left[ <\Phi ^2>_\phi -<\Phi
^2>_0\right] (x)=g\xi (x)  \label{lanleq2}
\end{equation}

Our goal is to show that any plane wave light field background will be
followed by a slowly varying echo. Since the light mass is nonvanishing
there is no loss of generality for our purpose if we assume the light field
is homogeneous in space and harmonic in time, i. e.,

\begin{equation}
\phi (t)=\phi _0\sin 2\omega t.  \label{bkg}
\end{equation}

The condition that the light four momentum lies below the branch point at $%
-k^2=4M^2$ translates into $\omega \leq M$.

To compute the nonperturbative noise kernel, we decompose the quantum heavy
fields propagating on the light background field into normal modes. The
amplitudes of each normal mode are complex, with 
\begin{equation}
\Phi _{-\vec k}=\Phi _{\vec k}^{\dagger }
\end{equation}
They obey the wave equation

\begin{equation}
\partial _t^2\Phi _{\vec k}+\Omega _k^2\Phi _{\vec k}=0
\end{equation}
where

\begin{equation}
\Omega _k^2=\vec k^2+M^2+g\phi (t)
\end{equation}
is the natural frequency of the $\vec k$th mode. Here we shall disregard the
possibility of $\Omega$ becoming imaginary through a large negative light
field, i.e., we assume $g\phi _0 \le M^2$. The strength of interaction
between the light and heavy fields is measured by

\begin{equation}
\kappa _k=\frac 1{2\Omega _k}\frac{d\Omega _k}{dt}=\frac 1{4\Omega _k^2}%
\frac{d\Omega _k^2}{dt}.
\end{equation}

We assume the heavy field is in the vacuum state at some initial time $t=0$.
Since it is a free field, Wick's theorem holds, and our problem is to relate
the field at abitrary times to the initial creation and destruction
operators. Of course, without knowing the explicit evolution law for the
light field, we cannot get the exact form, but have to find a suitable
approximation scheme. The general relationship we seek is

\begin{equation}
\Phi _k(t)=f_k(t)a_k(0)+f_k^{*}(t)a_{-k}^{\dagger }(0)  \label{secquant}
\end{equation}
where $f_k$ is the positive frequency mode associated to the IN particle
model \cite{nfsg}. It can be decomposed into instantaneous positive and
negative frequency parts as

\begin{equation}
f_k(t)=\frac 1{\sqrt{2\Omega _k}}\left[ \alpha _k(t)+\beta _k(t)\right]
\end{equation}

\subsection{Stochastic features near threshold}

Let us first consider the near threshold ($\omega \sim M$), weak field
regime, where $\Omega _k$ is essentially constant, and

\begin{equation}
\kappa _k\sim 2c_k\cos 2\omega t
\end{equation}
where 
\begin{equation}
c_k\sim \frac{\omega g\phi _0}{4\Omega _k^2}.  \label{ck}
\end{equation}

The Bogolubov coefficients $\alpha _k,\beta _k$ are calculated in Appendix A
to be 
\begin{equation}
\alpha _k(t)=\left( \frac{c_k}{2\gamma _k}\right) e^{-i\omega t}e^{\gamma
_kt}\left( 1+e^{i\delta _k}e^{-2\gamma _kt}\right) e^{-i\delta _k/2}
\label{alfabog}
\end{equation}

\begin{equation}
\beta _k(t)=\left( \frac{c_k}{2\gamma _k}\right) e^{i\omega t}e^{\gamma
_kt}\left( 1-e^{-2\gamma _kt}\right)  \label{betabog}
\end{equation}
where

\begin{equation}
e^{i\delta _k/2}=\left( \frac{\gamma _k}{c_k}\right) +i\left( \frac{\Omega
_k-\omega }{c_k}\right)  \label{deltabog}
\end{equation}

\begin{equation}
\gamma _k=\sqrt{c_k^2-(\Omega _k-\omega )^2}
\end{equation}

We observe that eqs. (\ref{alfabog}) and (\ref{betabog}) are formally valid
in the whole range of frequencies. However, outside the parametric resonance
regime, we have

\begin{equation}
\gamma _k\sim \pm i(\Omega _k-\omega )
\end{equation}
and both $\alpha _k$ and $\beta _k$ describe positive frequency oscillations
above the heavy threshold. We are interested here in the opposite case,
where three features stand out, namely, 1) the generation of the negative
frequency components described by $\beta _k$, which is the physical basis
for vacuum particle creation; 2) the exponential amplification due to
ongoing particle creation, and 3) the phase-locking of a whole range of
wavelengths at the resonance frequency $\omega $. As we shall now see, phase
locking allows the generation of a low frequency, inhomogeneus stochastic
field, which can be detected at the scale of the light sector. This is the
main physical indication of the new features of dissipation and fluctuation
below threshold we want to highlight in the context of effective field
theory.

In order to find the noise kernel, let us decompose the Heisenberg operator $%
\Phi ^2$ into a c-number, a diagonal (D) and a nondiagonal (ND) (in the
particle number basis) part

\begin{equation}
\Phi ^2=\left\langle \Phi ^2\right\rangle _\phi +\Phi _D^2+\Phi _{ND}^2 
\end{equation}
where the (D) and (ND) components are

\begin{equation}
\Phi _D^2=\int \frac{d^3k}{\left( 2\pi \right) ^3}\frac{d^3k^{\prime }}{%
\left( 2\pi \right) ^3}\;e^{i(k+k^{\prime })x}\left\{ f_k(t)f_{k^{\prime
}}^{*}(t)a_{-k^{\prime }}^{\dagger }a_k+f_k^{*}(t)f_{k^{\prime
}}(t)a_{-k}^{\dagger }a_{k^{\prime }}\right\}
\end{equation}

\begin{equation}
\Phi _{ND}^2=\int \frac{d^3k}{\left( 2\pi \right) ^3}\frac{d^3k^{\prime }}{%
\left( 2\pi \right) ^3}\;e^{i(k+k^{\prime })x}\left\{ f_k(t)f_{k^{\prime
}}(t)a_ka_{k^{\prime }}+f_k^{*}(t)f_{k^{\prime }}^{*}(t)a_{-k}^{\dagger
}a_{-k^{\prime }}^{\dagger }\right\}.
\end{equation}
Observe that

\begin{equation}
\left\langle \Phi _D^2\right\rangle _\phi =\left\langle \Phi
_{ND}^2\right\rangle _\phi =\left\langle \Phi _D^2\Phi _{ND}^2\right\rangle
_\phi =\left\langle \Phi _D^2\Phi _D^2\right\rangle _\phi \equiv 0 .
\end{equation}
Therefore,

\begin{eqnarray*}
N(x,x^{\prime }) &=&\left( \frac 18\right) \left\langle \left\{ \Phi
_{ND}^2(x),\Phi _{ND}^2(x^{\prime })\right\} \right\rangle _{\left\{ \phi
\right\} } \\
&=&\left( \frac 12\right) \int \frac{d^3k}{\left( 2\pi \right) ^3}\frac{%
d^3k^{\prime }}{\left( 2\pi \right) ^3}\;e^{i(k+k^{\prime })(x-x^{\prime })}%
{\rm Re}\left\{ f_k(t)f_{k^{\prime }}(t)f_k^{*}(t^{\prime })f_{k^{\prime
}}^{*}(t^{\prime })\right\}
\end{eqnarray*}

If no particle creation occurred, the noise kernel would contain frequencies
above threshold only. However, in the presence of frequency-locking and a
negative frequency part of the mode functions $f$, the noise kernel also
contains a steady component

\begin{equation}
N_S(x,x^{\prime })=\left( \frac 12\right) \int^{\prime }\frac{d^3k}{\left(
2\pi \right) ^3\Omega _k}\frac{d^3k^{\prime }}{\left( 2\pi \right) ^3\Omega
_{k^{\prime }}}\;e^{i(k+k^{\prime })(x-x^{\prime })}\Theta _{kk^{\prime
}}(t,t^{\prime }) 
\end{equation}
where the integral is restricted to those modes where $\gamma _k$ is real,
and

\begin{equation}
\Theta _{kk^{\prime }}(t,t^{\prime })={\rm Re}\left\{ \left( \alpha
_k(t)\beta _{k^{\prime }}(t)+\alpha _{k^{\prime }}(t)\beta _k(t)\right)
\left( \alpha _k(t^{\prime })\beta _{k^{\prime }}(t^{\prime })+\alpha
_{k^{\prime }}(t^{\prime })\beta _k(t^{\prime })\right) ^{*}\right\} . 
\end{equation}

It is important to notice that $\Theta $ is slowly varying not only with
respect to the heavy frequencies $\Omega $, but also with respect to the
locking frequency $\omega $. Of course we do not observe the noise kernel
directly, but only through its effect on the light field. However, since the
steady part of the stochastic source is slowly varying in space and time, to
first approximation it induces a stochastic light field $\phi_S$ which is
simply proportional to it

\begin{equation}
\phi _S\sim \left( \frac g{m^2}\right) \xi _S\qquad ;\qquad \left\langle
\phi _S\phi _S\right\rangle \sim \left( \frac g{m^2}\right) ^2N_S . 
\end{equation}
This is the echo we sought for. One can deduce the noise and its
auto-correlation in this way.

It is interesting to show the actual form of the noise kernel in the
opposite limits of very long and very short times, as we now do.

\subsubsection{Long time limit}

At long times, the correlation function is dominated by the very long
wavelength modes. We may thus approximate

\begin{equation}
c_k\sim c_0\sim \frac{\omega g\phi _0}{4M^2} 
\end{equation}

\begin{equation}
\gamma _k\sim \gamma _0-\frac{\sigma ^2k^2}2
\end{equation}
in the exponents, where

\begin{equation}
\gamma _0=\sqrt{c_0^2-\left( M-\omega \right) ^2}
\end{equation}

\begin{equation}
\sigma ^2=\frac 1{\gamma _0}\left[ 1-\frac \omega M+\frac{2\omega ^2g^2\phi
_0^2}{M^6}\right]
\end{equation}
neglecting $k^2$ elsewhere. Moreover, for $\gamma _0t\gg 1$ we neglect the
decaying modes, and extend the integral to all $k$ space. The result is

\begin{equation}
\alpha _k(t)\sim \left( \frac{c_0}{2\gamma _0}\right) e^{-i\omega
t}e^{\gamma _kt}e^{-i\delta _0/2} 
\end{equation}

\begin{equation}
\beta _k(t)=\left( \frac{c_0}{2\gamma _0}\right) e^{i\omega t}e^{\gamma _kt} 
\end{equation}

\begin{equation}
\alpha _k(t)\beta _{k^{\prime }}(t)+\alpha _{k^{\prime }}(t)\beta _k(t)\sim
2\left( \frac{c_0}{2\gamma _0}\right) ^2e^{(\gamma _k+\gamma _{k^{\prime
}})t}e^{-i\delta _0/2} 
\end{equation}

\begin{equation}
\Theta _{kk^{\prime }}(t,t^{\prime })=4\left( \frac{c_0}{2\gamma _0}\right)
^4e^{(\gamma _k+\gamma _{k^{\prime }})(t+t^{\prime })} 
\end{equation}

\begin{equation}
N_S(x,x^{\prime })=\left( \frac 1{8\left( 2\pi \right) ^6M^2}\right) \left( 
\frac{c_0}{\gamma _0}\right) ^4e^{2\gamma _0(t+t^{\prime })}\left[ \int
d^3k\;e^{ik(x-x^{\prime })}e^{-\sigma ^2k^2(t+t^{\prime })/2}\right] ^2 
\end{equation}

Performing the Gaussian integrals

\begin{equation}
N_S(x,x^{\prime })\sim \left( \frac{\omega g\phi _0}{4M^2}\right) ^4\frac{%
e^{2\gamma _0(t+t^{\prime })}}{\gamma _0^4M^2}\frac{e^{-(x-x^{\prime
})^2/\sigma ^2(t+t^{\prime }))}}{\left( 2\pi \sigma ^2(t+t^{\prime })\right)
^3}
\end{equation}
We observe that a large scale, inhomogeneous stochastic structure in the
light sector emerges from the back reaction of the created pairs of the
heavy field. This structure takes the form of domains where the field is
aligned, and the charactheristic size of these domains grows as the square
root of time.

\subsubsection{Short time limit}

In the opposite, very short time limit, we find

\begin{equation}
\alpha _k\approx e^{-i\omega t}  \label{alfastl}
\end{equation}

\begin{equation}
\beta _k(t)=c_kte^{i\omega t}  \label{betastl}
\end{equation}

\begin{equation}
\alpha _k(t)\beta _{k^{\prime }}(t)+\alpha _{k^{\prime }}(t)\beta _k(t)\sim
\left( c_k+c_{k^{\prime }}\right) t 
\end{equation}

\begin{equation}
\Theta _{kk^{\prime }}(t,t^{\prime })=\left( c_k+c_{k^{\prime }}\right)
^2tt^{\prime } 
\end{equation}

\begin{equation}
N_S(x,x^{\prime })=\left( \frac{tt^{\prime }}2\right) \int^{\prime }\frac{%
d^3k}{\left( 2\pi \right) ^3\Omega _k}\frac{d^3k^{\prime }}{\left( 2\pi
\right) ^3\Omega _{k^{\prime }}}\;e^{i(k+k^{\prime })(x-x^{\prime })}\left(
c_k+c_{k^{\prime }}\right) ^2 
\end{equation}
Approximately,

\begin{equation}
N_S(x,x^{\prime })\sim \left( \frac{c_0^2k_0^6}{2\pi ^4\Omega _0^2}\right)
tt^{\prime }f^2(k_0r)  \label{ns}
\end{equation}
where $r=\left| \vec x-\vec x^{\prime }\right| $,

\begin{equation}
f(u)=\left( \frac 1{u^3}\right) \left[ u\cos u-\sin u\right] 
\end{equation}
and $k_0$ marks the boundary of the resonant zone,

\begin{equation}
k_0\sim \sqrt{\left( \omega +c_0\right) ^2-M^2}. 
\end{equation}

As before, we should stress that the scale of the stochastic echo is much
lower than threshold. Even in the $\omega \rightarrow M$ limit, we find

\begin{equation}
k_0\sim \omega \sqrt{\frac{g\phi _0}{2M^2}}\ll \omega 
\end{equation}

\subsection{Stochastic behavior far below threshold}

Let us now consider the physically most relevant case, when the frequency of
the light background wave is far below the heavy threshold. As before, we
assume $\phi (t)=\phi _0\sin 2\omega t$, so that

\begin{equation}
\Omega _k\sim \Omega _{k0}+\delta \Omega _k 
\end{equation}

\begin{equation}
\delta \Omega _k\sim \frac{g\phi _0}{2\Omega _{k0}}\sin 2\omega t 
\end{equation}

\begin{equation}
\kappa _k=\frac 1{2\Omega _k}\frac{d\Omega _k}{dt}\sim 2c_k\cos 2\omega t 
\end{equation}
where

\begin{equation}
c_k=\frac{g\phi _0}{4\Omega _{k0}^2}\omega 
\end{equation}
We are interested in the case where $c_k\ll \omega $, and we assume

\begin{equation}
\frac{\Omega _{k0}}\omega =\left( 2N+1\right) \left( 1+\delta \right) 
\end{equation}
with $N\gg 1\gg \delta $.

As we show in Appendix B, the Bogolubov coefficients are given by

\begin{equation}
\alpha _k(t)=\left( \frac{C _k}{2\Gamma _k}\right) e^{\Gamma _kt}\left(
1+e^{i\Delta _k}e^{-2\Gamma _kt}\right) e^{-i\Delta _k/2}e^{-i\Theta _k}
\end{equation}

\begin{equation}
\beta _k(t)=\left( \frac{C _k}{2\Gamma _k}\right) e^{\Gamma _kt}\left(
1-e^{-2\Gamma _kt}\right) e^{i\Theta _k}
\end{equation}
where

\begin{equation}
e^{i\Delta _k/2}=\left( \frac 1{C _k}\right) \left\{ \Gamma _k+i\left[
\Omega _{k0}-(2N+1)\omega \right] \right\}
\end{equation}

\begin{equation}
\Gamma _k=\sqrt{C _k^2-\left[ \Omega _{k0}-(2N+1)\omega \right] ^2}
\end{equation}

\begin{equation}
\Theta _k=(2N+1)\omega t-\left( \frac{\Omega _kc_k}{\omega ^2}\right) \cos
2\omega t 
\end{equation}
and
\begin{equation}
C _k=c_k{\rm J}_{2N}\left( \frac{2\Omega _{k0}c_k}{\omega ^2}\right) 
\end{equation}
where J represents the usual Bessel function. When $N$ is large, the
asymptotics of Bessel functions yields \cite{couhil}

\begin{equation}
C _k\sim \left( \frac{c_k}{\sqrt{\pi N\tanh a}}\right) e^{-2N(a-\tanh a)} 
\end{equation}
where $\cosh a=\omega /2c_k(1+\delta )$, or, in short,

\begin{equation}
a\sim \ln \left( \frac{4\Omega _{k0}^2}{g\phi _0}\right).
\end{equation}

As expected, both the amplitude and the growth rate of the stochastic `echo'
are exponentially suppressed. In terms of the analysis of the previous
subsection, this case always falls in the ''short time'' limit. The
amplitude and growth rate, as well as the inverse size, of a stochastic
domain shall be given by $C_0.$ At truly low scales, the effect is extremely
feeble, but it builds up exponentially as we reach for the heavy threshold.
Since in a realistic situation this effect may be masked by
self-interactions, this exponential scale dependence may be essential to its
detectability, as one carries out measurements at successively higher
energies in this below-threshold region.

The exponential supression of particle creation and back reaction below the
threshold brings to mind the analogy with quantum tunneling phenomena, which
also depend exponentially on the height of the potential barrier. \footnote{%
We thank Diego Mazzitelli for this observation}. In both cases, though, these
quantitatively small effects become important because of their qualitative
impact on the physics of the system, and because no other perturbative
effects are there to mask them. It is striking in this regard because
tunneling dynamics is extremely sensitive to dissipation \cite{leggett}, and
therefore to the kind of phenomena we are discussing.

\subsection{Dissipation below threshold}

As discussed in the Introduction, a noisy theory should also be dissipative.
It is interesting then to conclude our treatment of fluctuations with a
brief account of dissipative phenomena at low energies.

Dissipation is associated with the nonperturbative deterministic part of
the equation of motion Eq. (\ref{lanleq2}), namely,

\begin{equation}
(\frac g2)\left[ <\Phi ^2>_\phi -<\Phi ^2>_0\right] (x)
\end{equation}
It is straightforward to show that

\begin{equation}
<\Phi ^2>_\phi =\int \frac{d^3k}{\left( 2\pi \right) ^32\Omega _k}\left\{
1+2\left| \beta _k\right| ^2+2{\rm Re}\left[ \alpha _k\beta _k^{*}\right]
\right\} 
\end{equation}
Neglecting the dependence of $\Omega _k$ on the light field, the vacuum
subtraction amounts to deleting the first term within brackets. 
The second term induces a deterministic, homogeneous shift in the low
frequency  light field. However, this effect is not associated to
dissipation, being a reversible vacuum polarization effect much alike the
Casimir energy between conducting plates \cite{birdav}.

It is the third term which depicts the truly dissipative effects.
At short times it amounts  to a viscous force

\begin{equation}
f=(\frac g2)\int^{\prime }\frac{d^3k}{\left( 2\pi \right) ^3\Omega _k}%
\left\{ c_kt\cos \omega t\right\} 
\end{equation}
(cf. Eq. (\ref{alfastl}, \ref{betastl})). The integral is restricted to
those modes where particle creation is effective. This force dissipates
energy from the oscillating light field, which must be provided by the
external agency sustaining the plane wave background.  The
energy dissipated per unit volume is

\begin{equation}
\delta \varepsilon =\int dt\;f\dot \phi \sim (\frac g2)\int^{\prime }\frac{%
d^3k}{\left( 2\pi \right) ^3\Omega _k}\left\{ \frac{\omega c_k\phi _0t^2}%
2\right\} \equiv \int \frac{d^3k}{\left( 2\pi \right) ^3}\Omega _k\left|
\beta _k\right| ^2
\end{equation}
(cf. Eqs. (\ref{bkg}, \ref{ck}, \ref{betastl})). This establishes the link
between dissipation and particle creation, and is essentially the same
result as obtained earlier via the perturbative approach (e.g., \cite{CH89}).

A fraction of the dissipated energy is returned to the system, degraded into
stochastic fluctuations. The stochastic source produces a total amount of
work per unit volume

\begin{equation}
\delta W\sim g\int dt\left\langle \xi \dot \phi _S\right\rangle \sim \left(
\frac gm\right) ^2\int dt\frac \partial {\partial t^{\prime
}}N_S(t,t^{\prime })\mid _{t^{\prime }\rightarrow t}\sim \left( \frac{%
g^2c_0^2k_0^6}{m^2\Omega _0^2}\right) t^2
\end{equation}
(cf. Eq. (\ref{ns})).

Under equilibrium conditions, the sum total of the dissipated energy equals
the total work done by the stochastic force integrated over time. This is 
a manifestation of a nonlinear fluctuation-dissipation relation. 
A precise statement of this involves the simultaneous consideration of 
several light modes, a task perhaps for future investigations.

\section{Discussions}

In this paper, we have presented a new way of looking at effective field
theories, bringing forth their intrinsically dissipative and stochastic
aspects. We have shown that dissipation and noise are generic features of
such theories, both below and above the energy threshold of the heavy mass
which defines their limit of applicability. As the threshold is crossed, the
character of the light theory does not change discontinuously, as commonly
believed, but is a continuous extension of what is already present below the
heavy scales. The stochastic features of the light theory (including the
build up of randomness and the breakdown of unitarity) though exceedingly
small, will manifest themselves at an exponentially increasing rate as the
energy is raised.

In the Introduction we have stressed the relevance of the observational
context in the definition of an open system, and in interpreting the
physical meaning of what is measured (e.g., appearance of dissipation and
fluctuations in an open system, but absence in a closed system) in the
restricted range of validity of the effective theory. In the same vein we
understand a light field as a representation of the full quantum field
observed at low energy. Standard texts tell us that this physical
field is obtained from the bare fields of the theory through the
renormalization process. However it is instructional to
reexamine the meaning of renormalization in an effective field theory from
the open-system viewpoint. Technically, renormalization means that the
effects of certain quantum degrees of freedom are added to the bare
quantities, and one regards these renormalized quantities as the actual
physically measureable ones (e.g., the mean energy in the Maxwell field
is added to the bare electron mass to make up its physical mass). In the
open-system viewpoint, which is closer to observation than the formally
complete yet unrealistic closed-system description (of all the constituents
at all energies), renormalization is a coarse graining operation:
certain `irrelevant' modes (in the above example, the virtual photons
surrounding an electron), considered as
the environment, are `slaved' (for definition see \cite{cddn}) to the
`relevant' modes of the particle, which constitute the (open) system, thus
enabling one to compute their mean effect on the relevant physics and come
up with an effective theory for the (open) system. Because this is an
essentially statistical operation, it carries with it the well known
statistical consequences: first, there is a gap between the mean value of a
system mode and the actual value which includes the backreaction of the
`irrelevant' modes, and for this difference the system will be subject
to a random source from the environment. Second, the approximations on which
the slaving procedure is
based (for example, to compute the quantum averages of the environment
variables it is often necessary to ignore or to downplay the backreaction of
these modes on the system variables) lose accuracy as the influence of the
`irrelevant' sector becomes large, as is the case when the fluctuations
become significant, their coupling becomes strong, or generally in the
long-time limit.

If one intends to have the light field represent the physical field, in a
strict sense, the set up of the effective field theory should include a
detailed account of the observational context, or at least of the
renormalization procedure involved. Presumably, there would be
transformation rules to translate the results from different renormalization
prescriptions, and these could eventually take the form of renormalization
group equations \cite{cdj,wetterich}. However, in the presence of a sizable
gap between a light and a heavy scale, as in the case studied here,
sensible prescriptions will label most of the heavy field modes as
environment, and most of the light modes as system. Thus we have adopted
in the above the
somewhat simplistic view of treating renormalization as the dressing of the
light fields by the heavy quantum fluctuations. One shortcoming of this
assumption is that, for example, if the light fields self interact, this
prescription will not eliminate all infinities from the theory.

In the region where the system and environment get progressively entangled,
the system dynamics will acquire a stochastic component, and become
dissipative. An arrow of time will also emerge in the effective theory. It
is of interest to develop a renormalization group theory for dissipative
systems. Some of the traditional concepts would need a newer and broader
interpretation. The breakdown of an effective theory in the threshold region
is theoretically related to the cross-over behavior in critical phenomena
studied in depth by O'Connor and Stephens \cite{cdj}. Their
observation on how the relevant degrees of freedom of a physical theory are
dependent on the scales at which the theory is probed will be useful for the
construction of open systems which are sensitive to the energy and
observation scales. These are important questions at the foundation of
statistical mechanics and field theory which we hope to probe into.

The ideas presented in this paper can lead to several directions of further
development. At a basic level, there is the question about the fundamental
nature of any realistic physical system described by quantum field theories.
It is the view of the authors that in Nature there is no irreducibly
`fundamental' theories in the absolute sense, just as the existence of an
absolute closed system is more in the hypothetical rather than the physical
realm. (Even for the Universe, it is a closed system only in the ontological
rather than the physical sense.) All realistic theories describing open
systems are to varying degrees noisy and dissipative. They are depicted by
stochastic rather than strictly deterministic equations. Only when noise and
dissipation are small can one describe in approximate terms the system by
the usual tenets (e.g., effective action) of unitary field theory. The
criterion of validity of an effective field theory is derived in this paper.
In terms of structures, we also think that there are no irreducibly
elemental theories or constituents in the absolute sense. The presence of
noise, albeit in small amounts, points to the presence of a deeper layer of
structure. (To give a historical example, brownian motion marks the boundary
between hydrodynamics and many body theory, as it discloses the graininess
of a seemingly continuous fluid). Indeed this point of view can be used to
guide the probing into possible deeper and unknown layers of structures from
a better known, lower energy domain. The case of gravity, from the better
known semiclassical regime to the unknown quantum regime, was what motivated
us into examining the general properties of quantum open systems and
effective theories in the first place \cite{HuPhysica}. Noise and
fluctuations could then in this sense serve as a trace detector which allows
us to obtain a glimpse of the deeper structures.

There are also many physical situations where the mechanisms of fluctuation
generation and structure growth described in this paper could be put to
practical use. A particularly fertile ground is the physics of the Early
Universe, modeled by a theory of massless fields (gravitational, neutrinos
and gauge bosons) interacting with heavy fields such as electrons, quarks,
and cold dark matter candidates. The gravitational background will create
particles of the heavy fields (while neutrinos and gauge bosons are shielded
by conformal invariance), which in turn will react on the light fields,
resulting in the generation of primordial gravitational fluctuations \cite
{qfsf} and gauge fields. Closer to home, the theory of heavy ion collisions
also presents a situation where a color field background interacts with the
massive quark fields, resulting in the formation of a quark-gluon plasma,
which could be investigated using the framework of this paper \cite{hic}. We
hope to report on the result of this and related research in later
publications.\\

\noindent {\bf Acknowledgement}\\It is a pleasure to acknowledge numerous
discussions with our colleagues Denjoe O'Connor, Juan Pablo Paz, Chris
Stephens. Special thanks go to Diego Mazzitelli for his comments on an
earlier draft of this paper. Part of this work was done while both of us
enjoyed the hospitality of Professor Enric Verdaguer during our visit to
the Universidad de Barcelona. This research is supported
in part by the US National Science Foundation under grant PHY94-21849 and
INT95-09847, and by the University of Buenos Aires, CONICET, Fundaci\'on
Antorchas (Argentina) and the Commission for the European Communities under
contract $CI1*-CT94-0004$.

\newpage

\section{Appendices}

\subsection*{A. Derivation of $\alpha $ and $\beta $}

Our method can be described as a translation into Hamilton - Jacobi language
of the classical averaging method, as found in the textbooks by Landau and
Lifshitz \cite{lanlif} and Bogolubov and Mitropolsky \cite{bogmit} (it is
close to the methods used by \cite{stabra}.) We shall borrow some tools of
classical mechanics to approach our problem. 
The mode equation follows from the Hamiltonian

\begin{equation}
H=P_{-k}P_k+\Omega ^2\Phi _{-k}\Phi _k
\end{equation}
where $\Phi _k$ and $\Phi _{-k}$ are independent canonical variables, and $%
P_k$, $P_{-k}$ their conjugate momenta, respectively. We introduce creation
and destruction operators through

\begin{equation}
\Phi _k=\frac 1{\sqrt{2\Omega }}\left[ a_k+a_{-k}^{*}\right] 
\end{equation}

\begin{equation}
P_k=i\sqrt{\frac \Omega 2}\left[ a_k^{*}-a_{-k}\right]  \label{crdest}
\end{equation}
We adopt the destruction operators as new canonical variables, with
conjugated momenta

\begin{equation}
p_{\pm k}=ia_{\pm k}^{*}
\end{equation}

The Hamiltonian expressed in terms of these new variables is

\begin{equation}
K=-i\Omega \left[ a_kp_k+a_{-k}p_{-k}\right] -i\kappa \left[
a_ka_{-k}+p_kp_{-k}\right]
\end{equation}
where, as before,

\begin{equation}
\kappa =\frac 1{2\Omega }\frac{d\Omega }{dt}
\end{equation}

The Bogolubov transformation linking the destruction and creation operators
at time $t$ with those at time $t=0$ is given by

\begin{equation}
a_k(t)=\alpha (t)a_k(0)+\beta ^{*}(t)a_{-k}^{\dagger }(0)
\end{equation}

\begin{equation}
a_k^{\dagger }(t)=\alpha ^{*}(t)a_k^{\dagger }(0)+\beta (t)a_{-k}(0)
\label{bogotr}
\end{equation}
where the coefficients satisfy the Wronkian condition

\begin{equation}
\left| \alpha \right| ^2-\left| \beta \right| ^2=1
\end{equation}

In terms of the canonical variables, the Bogolubov transformation takes the
form

\begin{equation}
a_k(t)=\alpha (t)a_k(0)-i\beta ^{*}(t)p_{-k}(0) 
\end{equation}

\begin{equation}
p_k(t)=\alpha ^{*}(t)p_k(0)+i\beta (t)a_{-k}(0)
\end{equation}
This is a canonical transformation with generating functional

\begin{equation}
S=G(t)\left[ a_kp_k(0)+a_{-k}p_{-k}(0)\right]
+F(t)a_ka_{-k}+E(t)p_k(0)p_{-k}(0)
\end{equation}
where (from now on, we shall occasionally omit the $k$ subindices, to
simplify the appearance of our formulae)

\begin{equation}
G=\frac 1\alpha, ~~ E=\frac{i\beta ^{*}}\alpha, ~~ F=\frac{i\beta }\alpha
\end{equation}
$S$ satisfies the Hamilton-Jacobi equation

\begin{equation}
-i\Omega \left[ a_k\frac{\partial S}{\partial a_k}+a_{-k}\frac{\partial S}{%
\partial a_{-k}}\right] -i\kappa \left[ a_ka_{-k}+\frac{\partial S}{\partial
a_k}\frac{\partial S}{\partial a_{-k}}\right] +\frac{\partial S}{\partial t}%
=0
\end{equation}
Therefore,

\begin{equation}
\frac{dE}{dt}-i\kappa G^2=0
\end{equation}

\begin{equation}
\frac{dG}{dt}-i\left[ \Omega +\kappa F\right] G=0  \label{geq}
\end{equation}

\begin{equation}
\frac{dF}{dt}-2i\Omega F-i\kappa \left[ 1+F^2\right] =0  \label{feq}
\end{equation}

Short of an exact solution, the conventional appproach to solving this
equation would be to expand in powers of $\kappa $. This leads to the usual
adiabatic approximation \cite{ParFul,ZelSta}, since $\kappa \sim O(g)$,
which is precisely what we should avoid for the present purpose.

As before, let us assume $\Omega $ is essentially constant, and

\begin{equation}
\kappa \sim 2c\cos 2\omega t
\end{equation}
where 
\begin{equation}
c\sim \frac{\omega g\phi _0}{4\Omega _k^2}
\end{equation}
with $\omega \leq \Omega _k$. The idea is to retain only the most resonant
terms in Eq. (\ref{feq}); namely, we write

\begin{equation}
\frac{dF}{dt}-2i\Omega _kF-ic\left[ e^{2i\omega t}+e^{-2i\omega t}F^2\right]
=0
\end{equation}
This equation allows a solution of the form

\begin{equation}
F=\frac ice^{2i\omega t}\frac{\dot u}u  \label{ansatz}
\end{equation}
where $u$ satisfies the ordinary equation

\begin{equation}
\frac{d^2u }{dt^2}-2i(\Omega -\omega )\frac{du }{dt}-c^2u =0.
\end{equation}

The solutions are

\begin{equation}
u _{\pm }\sim e^{\pm \gamma t}e^{i(\Omega -\omega )t}
\end{equation}
where

\begin{equation}
\gamma =\sqrt{c^2-(\Omega -\omega )^2}
\end{equation}
The case of interest to us is when $\gamma $ is real.

To find $\alpha $ we integrate Eq. (\ref{geq}) under the approximation

\begin{equation}
\kappa F\sim ce^{-2i\omega t}F\equiv i\frac{\dot u}u
\end{equation}
that is, we keep only the slowly varying term. The integration is then
trivial, and we get

\begin{equation}
G=\frac{e^{i\Omega t}}u\quad ;\alpha =ue^{-i\Omega t} 
\end{equation}

Given $F$ and $\alpha $, finding $\beta $ is a matter of algebra

\begin{equation}
\beta =-i\alpha F\equiv \frac{\dot u}ce^{i(2\omega -\Omega )t} 
\end{equation}
We thus find the boundary conditions $u(0)=1$, $\dot u(0)=0$. The solution is

\begin{equation}
u=\left( \frac c{2\gamma }\right) \left[ e^{-i\delta /2}e^{\gamma
t}+e^{i\delta /2}e^{-\gamma t}\right] e^{i(\Omega -\omega )t} 
\end{equation}
leading to Eqs. (\ref{alfabog}), (\ref{betabog}), (\ref{deltabog}) as given
in the main text.

\subsection*{B. Particle Creation Far Below Threshold}

When the frequency $\omega $ of the normal modes of the light field is far
below $M$, the above analysis is valid up to Eq. (\ref{feq}), but care must
be taken to identify the resonant terms. Let us decompose the frequency $%
\Omega $ into constant and fluctuating parts

\begin{equation}
\Omega =\Omega _0+\delta \Omega
\end{equation}
Then, from

\begin{equation}
\kappa =\frac 1{2\Omega }\frac{d\Omega }{dt}\sim 2c\cos 2\omega t
\end{equation}
we get

\begin{equation}
\delta \Omega \sim \left( \frac{2\Omega _0c}\omega \right) \sin 2\omega t
\end{equation}

Let us assume

\begin{equation}
\frac{\Omega _0}\omega \equiv (2N+1)\left( 1+\delta \right) ,
\end{equation}
where $N$ is an integer and $\delta \ll 1$. Then resonance occurs at the
frequency $(2N+1)\omega $.

Instead of Eq. (\ref{ansatz}), we now try

\begin{equation}
F=\left( \frac{i(-1)^Ne^{2i\Theta }} C \right) \left[ \frac{\dot U}U\right] 
\end{equation}
where

\begin{equation}
\Theta =(2N+1)\omega t-\left( \frac{\Omega c}{\omega ^2}\right) \cos 2\omega
t 
\end{equation}

Expanding the exponential as a Fourier series, and keeping only the resonant
term, we find

\begin{equation}
\kappa e^{\pm 2i\Theta }\sim (-1)^Nc{\rm J}_{2N}\left( \frac{2\Omega c}{%
\omega ^2}\right) \equiv (-1)^N C 
\end{equation}
The equation for $U$ reads

\begin{equation}
\ddot U-2i\left[ \Omega -\left( 2N+1\right) \omega \right] \dot U-C^2U=0 
\end{equation}
>From here on, the argument exactly reproduces the previous case, leading to
the results reported in the text.


\end{document}